\documentclass[12pt]{article}
\usepackage{amsmath,amssymb}
\numberwithin{equation}{section}
\title{Chiral Algebras of $(0,2)$ Sigma Models:\\
Beyond Perturbation Theory --- II \\[12pt]}

\author{\normalsize Meng-Chwan~Tan\footnote{On leave of absence from the National University of Singapore.} \\
\normalsize\it
School of Natural Sciences, Institute for Advanced Study \\
\normalsize\it
Princeton, NJ 08540, USA  \\[2pt]
\normalsize and \\[2pt]
\normalsize Junya Yagi \\
\normalsize\it
Department of Physics and Astronomy, Rutgers University \\
\normalsize\it
Piscataway, NJ 08855, USA}

\date{}

\makeatletter
\renewcommand\section{\@startsection {section}{1}{\z@}%
                                   {-3.5ex \@plus -1ex \@minus -.2ex}%
                                   {2.3ex \@plus.2ex}%
                                   {\normalfont\Large\bfseries\boldmath}}
\renewcommand\subsection{\@startsection{subsection}{2}{\z@}%
                                     {-3.25ex\@plus -1ex \@minus -.2ex}%
                                     {1.5ex \@plus .2ex}%
                                     {\normalfont\large\bfseries\boldmath}}
\renewcommand\subsubsection{\@startsection{subsubsection}{3}{\z@}%
                                     {-3.25ex\@plus -1ex \@minus -.2ex}%
                                     {1.5ex \@plus .2ex}%
                                     {\normalfont\normalsize\bfseries\boldmath}}
\renewcommand\paragraph{\@startsection{paragraph}{4}{\z@}%
                                    {3.25ex \@plus1ex \@minus.2ex}%
                                    {-1em}%
                                    {\normalfont\normalsize\bfseries\boldmath}}
\renewcommand\subparagraph{\@startsection{subparagraph}{5}{\parindent}%
                                       {3.25ex \@plus1ex \@minus .2ex}%
                                       {-1em}%
                                      {\normalfont\normalsize\bfseries\boldmath}}
\makeatother

\newcommand{\del}{\partial}

\newcommand{\TXb}{\overline{TX}}
\newcommand{\CP}{\mathbb{CP}}

\newcommand{\ket}[1]{|#1\rangle}
\newcommand{\bra}[1]{\langle #1|}

\newcommand{\vev}[1]{\langle #1 \rangle}
\newcommand{\bigvev}[1]{\bigl\langle #1 \bigr\rangle}

\newcommand{\Sym}{\mathop{\mathrm{Sym}}\nolimits}

\newcommand{\iso}{\cong}

\newcommand{\Z}{\mathbb{Z}}

\newcommand{\R}{\mathbb{R}}

\renewcommand{\sl}{\mathit{sl}}
\newcommand{\SL}{\mathit{SL}}

\let\nc\newcommand
\let\ovl\overline
\let\td\tilde
\let\wtd\widetilde
\let\wht\widehat
\let\mcl\mathcal

\newcommand{\ab}{{\bar{a}}}
\newcommand{\bb}{{\bar{b}}}

\newcommand{\hb}{{\bar{h}}}
\newcommand{\ib}{{\bar{\imath}}}
\newcommand{\jb}{{\bar{\jmath}}}
\newcommand{\kb}{{\bar{k}}}

\newcommand{\ub}{{\bar{u}}}

\newcommand{\wb}{{\bar{w}}}

\newcommand{\zb}{{\bar{z}}}

\nc{\Ab}{{\ovl{A}}} \nc{\At}{{\wtd{A}}} \nc{\Ah}{{\wht{A}}}
\nc{\Bb}{{\ovl{B}}} \nc{\Bt}{{\wtd{B}}} \nc{\Bh}{{\wht{B}}}
\nc{\Cb}{{\ovl{C}}} \nc{\Ct}{{\wtd{C}}} \nc{\Ch}{{\wht{C}}}
\nc{\Db}{{\ovl{D}}} \nc{\Dt}{{\wtd{D}}} \nc{\Dh}{{\wht{D}}}
\nc{\Eb}{{\ovl{E}}} \nc{\Et}{{\wtd{E}}} \nc{\Eh}{{\wht{E}}}
\nc{\Fb}{{\ovl{F}}} \nc{\Ft}{{\wtd{F}}} \nc{\Fh}{{\wht{F}}}
\nc{\Gb}{{\ovl{G}}} \nc{\Gt}{{\wtd{G}}} \nc{\Gh}{{\wht{G}}}
\nc{\Hb}{{\ovl{H}}} \nc{\Ht}{{\wtd{H}}} \nc{\Hh}{{\wht{H}}}
\nc{\Ib}{{\ovl{I}}} \nc{\It}{{\wtd{I}}} \nc{\Ih}{{\wht{I}}}
\nc{\Jb}{{\ovl{J}}} \nc{\Jt}{{\wtd{J}}} \nc{\Jh}{{\wht{J}}}
\nc{\Kb}{\smash{{\ovl{K}}}} \nc{\Kt}{{\wtd{K}}} \nc{\Kh}{{\wht{K}}}
\nc{\Lb}{{\ovl{L}}} \nc{\Lt}{{\wtd{L}}} \nc{\Lh}{{\wht{L}}}
\nc{\Mb}{{\ovl{M}}} \nc{\Mt}{{\wtd{M}}} \nc{\Mh}{{\wht{M}}}
\nc{\Nb}{{\ovl{N}}} \nc{\Nt}{{\wtd{N}}} \nc{\Nh}{{\wht{N}}}
\nc{\Ob}{{\ovl{O}}} \nc{\Ot}{{\wtd{O}}} \nc{\Oh}{{\wht{O}}}
\nc{\Pb}{{\ovl{P}}} \nc{\Pt}{{\wtd{P}}} \nc{\Ph}{{\wht{P}}}
\nc{\Qb}{{\ovl{Q}}} \nc{\Qt}{{\wtd{Q}}} \nc{\Qh}{{\wht{Q}}}
\nc{\Rb}{{\ovl{R}}} \nc{\Rt}{{\wtd{R}}} \nc{\Rh}{{\wht{R}}}
\nc{\Sb}{{\ovl{S}}} \nc{\St}{{\wtd{S}}} \nc{\Sh}{{\wht{S}}}
\nc{\Tb}{{\ovl{T}}} \nc{\Tt}{{\wtd{T}}} \nc{\Th}{{\wht{T}}}
\nc{\Ub}{{\ovl{U}}} \nc{\Ut}{{\wtd{U}}} \nc{\Uh}{{\wht{U}}}
\nc{\Vb}{{\ovl{V}}} \nc{\Vt}{{\wtd{V}}} \nc{\Vh}{{\wht{V}}}
\nc{\Wb}{{\ovl{W}}} \nc{\Wt}{{\wtd{W}}} \nc{\Wh}{{\wht{W}}}
\nc{\Xb}{{\ovl{X}}} \nc{\Xt}{{\wtd{X}}} \nc{\Xh}{{\wht{X}}}
\nc{\Yb}{{\ovl{Y}}} \nc{\Yt}{{\wtd{Y}}} \nc{\Yh}{{\wht{Y}}}
\nc{\Zb}{{\ovl{Z}}} \nc{\Zt}{{\wtd{Z}}} \nc{\Zh}{{\wht{Z}}}

\nc{\CA}{{\mcl{A}}} \nc{\CAb}{{\ovl{\CA}}} \nc{\CAt}{{\wtd{\CA}}} \nc{\CAh}{{\wht{\CA}}}
\nc{\CB}{{\mcl{B}}} \nc{\CBb}{{\ovl{\CB}}} \nc{\CBt}{{\wtd{\CB}}} \nc{\CBh}{{\wht{\CB}}}
\nc{\CC}{{\mcl{C}}} \nc{\CCb}{{\ovl{\CC}}} \nc{\CCt}{{\wtd{\CC}}} \nc{\CCh}{{\wht{\CC}}}
\nc{\CD}{{\mcl{D}}} \nc{\CDb}{{\ovl{\CD}}} \nc{\CDt}{{\wtd{\CD}}} \nc{\CDh}{{\wht{\CD}}}
\nc{\CE}{{\mcl{E}}} \nc{\CEb}{{\ovl{\CE}}} \nc{\CEt}{{\wtd{\CE}}} \nc{\CEh}{{\wht{\CE}}}
\nc{\CF}{{\mcl{F}}} \nc{\CFb}{{\ovl{\CF}}} \nc{\CFt}{{\wtd{\CF}}} \nc{\CFh}{{\wht{\CF}}}
\nc{\CG}{{\mcl{G}}} \nc{\CGb}{{\ovl{\CG}}} \nc{\CGt}{{\wtd{\CG}}} \nc{\CGh}{{\wht{\CG}}}
\nc{\CH}{{\mcl{H}}} \nc{\CHb}{{\ovl{\CH}}} \nc{\CHt}{{\wtd{\CH}}} \nc{\CHh}{{\wht{\CH}}}
\nc{\CI}{{\mcl{I}}} \nc{\CIb}{{\ovl{\CI}}} \nc{\CIt}{{\wtd{\CI}}} \nc{\CIh}{{\wht{\CI}}}
\nc{\CJ}{{\mcl{J}}} \nc{\CJb}{{\ovl{\CJ}}} \nc{\CJt}{{\wtd{\CJ}}} \nc{\CJh}{{\wht{\CJ}}}
\nc{\CK}{{\mcl{K}}} \nc{\CKb}{{\ovl{\CK}}} \nc{\CKt}{{\wtd{\CK}}} \nc{\CKh}{{\wht{\CK}}}
\nc{\CL}{{\mcl{L}}} \nc{\CLb}{{\ovl{\CL}}} \nc{\CLt}{{\wtd{\CL}}} \nc{\CLh}{{\wht{\CL}}}
\nc{\CM}{{\mcl{M}}} \nc{\CMb}{{\ovl{\CM}}} \nc{\CMt}{{\wtd{\CM}}} \nc{\CMh}{{\wht{\CM}}}
\nc{\CN}{{\mcl{N}}} \nc{\CNb}{{\ovl{\CN}}} \nc{\CNt}{{\wtd{\CN}}} \nc{\CNh}{{\wht{\CN}}}
\nc{\CO}{{\mcl{O}}} \nc{\COb}{{\ovl{\CO}}} \nc{\COt}{{\wtd{\CO}}} \nc{\COh}{{\wht{\CO}}}
\nc{\CQ}{{\mcl{Q}}} \nc{\CQb}{{\ovl{\CQ}}} \nc{\CQt}{{\wtd{\CQ}}} \nc{\CQh}{{\wht{\CQ}}}
\nc{\CR}{{\mcl{R}}} \nc{\CRb}{{\ovl{\CR}}} \nc{\CRt}{{\wtd{\CR}}} \nc{\CRh}{{\wht{\CR}}}
\nc{\CS}{{\mcl{S}}} \nc{\CSb}{{\ovl{\CS}}} \nc{\CSt}{{\wtd{\CS}}} \nc{\CSh}{{\wht{\CS}}}
\nc{\CT}{{\mcl{T}}} \nc{\CTb}{{\ovl{\CT}}} \nc{\CTt}{{\wtd{\CT}}} \nc{\CTh}{{\wht{\CT}}}
\nc{\CU}{{\mcl{U}}} \nc{\CUb}{{\ovl{\CU}}} \nc{\CUt}{{\wtd{\CU}}} \nc{\CUh}{{\wht{\CU}}}
\nc{\CV}{{\mcl{V}}} \nc{\CVb}{{\ovl{\CV}}} \nc{\CVt}{{\wtd{\CV}}} \nc{\CVh}{{\wht{\CV}}}
\nc{\CW}{{\mcl{W}}} \nc{\CWb}{{\ovl{\CW}}} \nc{\CWt}{{\wtd{\CW}}} \nc{\CWh}{{\wht{\CW}}}
\nc{\CX}{{\mcl{X}}} \nc{\CXb}{{\ovl{\CX}}} \nc{\CXt}{{\wtd{\CX}}} \nc{\CXh}{{\wht{\CX}}}
\nc{\CY}{{\mcl{Y}}} \nc{\CYb}{{\ovl{\CY}}} \nc{\CYt}{{\wtd{\CY}}} \nc{\CYh}{{\wht{\CY}}}
\nc{\CZ}{{\mcl{Z}}} \nc{\CZb}{{\ovl{\CZ}}} \nc{\CZt}{{\wtd{\CZ}}} \nc{\CZh}{{\wht{\CZ}}}

\let\eps\epsilon
\let\ups\upsilon
\let\veps\varepsilon
\let\vtht\vartheta
\let\vsgm\varsigma
\let\vphi\varphi
\let\vrho\varrho

\nc{\alphab}{{\bar{\alpha}}} \nc{\alphat}{{\td{\alpha}}} \nc{\alphah}{{\hat{\alpha}}}
\nc{\betab}{{\bar{\beta}}}   \nc{\betat}{{\td{\beta}}}   \nc{\betah}{{\hat{\beta}}} 
\nc{\gammab}{{\bar{\gamma}}} \nc{\gammat}{{\td{\gamma}}} \nc{\gammah}{{\hat{\gamma}}} 
\nc{\deltab}{{\bar{\delta}}} \nc{\deltat}{{\td{\delta}}} \nc{\deltah}{{\hat{\delta}}} 
\nc{\epsilonb}{{\bar{\eps}}} \nc{\epsilont}{{\td{\eps}}} \nc{\epsilonh}{{\hat{\eps}}} 
\nc{\vepsb}{{\bar{\veps}}}   \nc{\vepst}{{\td{\veps}}}   \nc{\vepsh}{{\hat{\veps}}} 
\nc{\zetab}{{\bar{\zeta}}}   \nc{\zetat}{{\td{\zeta}}}   \nc{\zetah}{{\hat{\zeta}}} 
\nc{\etab}{{\bar{\eta}}}     \nc{\etat}{{\td{\eta}}}     \nc{\etah}{{\hat{\eta}}} 
\nc{\thetab}{{\bar{\theta}}} \nc{\thetat}{{\td{\theta}}} \nc{\thetah}{{\hat{\theta}}} 
\nc{\vthetab}{{\bar{\vtht}}} \nc{\vthetat}{{\td{\vtht}}} \nc{\vthetah}{{\hat{\vtht}}} 
\nc{\iotab}{{\bar{\iota}}}   \nc{\iotat}{{\td{\iota}}}   \nc{\iotah}{{\hat{\iota}}} 
\nc{\kappab}{{\bar{\kappa}}} \nc{\kappat}{{\td{\kappa}}} \nc{\kappah}{{\hat{\kappa}}} 
\nc{\lmdb}{{\bar{\lmd}}}     \nc{\lmdt}{{\td{\lmd}}}     \nc{\lmdh}{{\hat{\lmd}}} 
\nc{\mub}{{\bar{\mu}}}       \nc{\mut}{{\td{\mu}}}       \nc{\muh}{{\hat{\mu}}} 
\nc{\nub}{{\bar{\nu}}}       \nc{\nut}{{\td{\nu}}}       \nc{\nuh}{{\hat{\nu}}} 
\nc{\xib}{{\bar{\xi}}}       \nc{\xit}{{\td{\xi}}}       \nc{\xih}{{\hat{\xi}}} 
\nc{\pib}{{\bar{\pi}}}       \nc{\pit}{{\td{\pi}}}       \nc{\pih}{{\hat{\pi}}} 
\nc{\vpib}{{\bar{\vpi}}}     \nc{\vpit}{{\td{\vpi}}}     \nc{\vpih}{{\hat{\vpi}}} 
\nc{\rhob}{{\bar{\rho}}}     \nc{\rhot}{{\td{\rho}}}     \nc{\rhoh}{{\hat{\rho}}} 
\nc{\vrhob}{{\bar{\vrho}}}   \nc{\vrhot}{{\td{\vrho}}}   \nc{\vrhoh}{{\hat{\vrho}}} 
\nc{\sigmab}{{\bar{\sigma}}} \nc{\sigmat}{{\td{\sigma}}} \nc{\sigmah}{{\hat{\sigma}}} 
\nc{\vsigmab}{{\bar{\vsgm}}} \nc{\vsigmat}{{\td{\vsgm}}} \nc{\vsigmah}{{\hat{\vsgm}}} 
\nc{\taub}{{\bar{\tau}}}     \nc{\taut}{{\td{\tau}}}     \nc{\tauh}{{\hat{\tau}}} 
\nc{\upsilonb}{{\bar{\ups}}} \nc{\upsilont}{{\td{\ups}}} \nc{\upsilonh}{{\hat{\ups}}} 
\nc{\phib}{{\bar{\phi}}}     \nc{\phit}{{\td{\phi}}}     \nc{\phih}{{\hat{\phi}}} 
\nc{\vphib}{{\bar{\vphi}}}   \nc{\vphit}{{\td{\vphi}}}   \nc{\vphih}{{\hat{\vphi}}} 
\nc{\chib}{{\bar{\chi}}}     \nc{\chit}{{\td{\chi}}}     \nc{\chih}{{\hat{\chi}}} 
\nc{\psib}{{\bar{\psi}}}     \nc{\psit}{{\td{\psi}}}     \nc{\psih}{{\hat{\psi}}} 
\nc{\omegab}{{\bar{\omega}}} \nc{\omegat}{{\td{\omega}}} \nc{\omegah}{{\hat{\omega}}} 

\nc{\Gammab}{{\ovl{\Gamma}}}     \nc{\Gammat}{{\wtd{\Gamma}}}     \nc{\Gammah}{{\wht{\Gamma}}}
\nc{\Deltab}{{\ovl{\Delta}}}     \nc{\Deltat}{{\wtd{\Delta}}}     \nc{\Deltah}{{\wht{\Delta}}}
\nc{\Thetab}{{\ovl{\Theta}}}     \nc{\Thetat}{{\wtd{\Theta}}}     \nc{\Thetah}{{\wht{\Theta}}}
\nc{\Lambdab}{{\ovl{\Lambda}}}   \nc{\Lambdat}{{\wtd{\Lambda}}}   \nc{\Lambdah}{{\wht{\Lambda}}}
\nc{\Xib}{{\ovl{\Xi}}}           \nc{\Xit}{{\wtd{\Xi}}}           \nc{\Xih}{{\wht{\Xi}}}
\nc{\Pib}{{\ovl{\Pi}}}           \nc{\Pit}{{\wtd{\Pi}}}           \nc{\Pih}{{\wht{\Pi}}}
\nc{\Sigmab}{{\ovl{\Sigma}}}     \nc{\Sigmat}{{\wtd{\Sigma}}}     \nc{\Sigmah}{{\wht{\Sigma}}}
\nc{\Upsilonb}{{\ovl{\Upsilon}}} \nc{\Upsilont}{{\wtd{\Upsilon}}} \nc{\Upsilonh}{{\wht{\Upsilon}}}
\nc{\Phib}{{\ovl{\Phi}}}         \nc{\Phit}{{\wtd{\Phi}}}         \nc{\Phih}{{\wht{\Phi}}}
\nc{\Psib}{{\ovl{\Psi}}}         \nc{\Psit}{{\wtd{\Psi}}}         \nc{\Psih}{{\wht{\Psi}}}
\nc{\Omegab}{{\ovl{\Omega}}}     \nc{\Omegat}{{\wtd{\Omega}}}     \nc{\Omegah}{{\wht{\Omega}}}

\begin{document}

\maketitle

\begin{abstract}
We extend our analysis in [arXiv:0801.4782] and show that the chiral algebras of $(0,2)$ sigma models are totally trivialized by worldsheet instantons for all complete flag manifolds of compact semisimple Lie groups.  Consequently, supersymmetry is spontaneously broken.  Our results verify Stolz's idea \cite{S} that there are no harmonic spinors on the loop spaces of these flag manifolds.  Moreover, they also imply that the kernels of certain twisted Dirac operators on these spaces will be empty under a ``quantum'' deformation of their geometries.
\end{abstract}

\vfill

\pagebreak

\section{Introduction}

The purpose of this paper is to explain how the results of \cite{TY} about the chiral algebra of a certain $\CN = (0,2)$ sigma model  generalize to a larger class of target spaces.  To begin, let us recall some of the basic facts about the chiral algebras of $(0,2)$ models.  We refer the reader to \cite{TY, W1} for more details.

Consider a $(0,2)$ model with worldsheet $\Sigma$ and target space $X$.  The twisted version of the theory possesses a global fermionic symmetry generated by the scalar supercharge~$Q$.  The basic property of $Q$ is that it is nilpotent: $Q^2 = 0$.  One can then immediately construct two $Q$-cohomology groups, namely, the cohomology of operators and the cohomology of states.  Let us focus on the cohomology of \emph{local} operators.  This is an infinite-dimensional space, graded by the right-moving $R$ charge and perturbatively, also by the scaling dimension.  This space of local operators has a natural ring structure defined by the OPE.  Unlike the $\CN = (2,2)$ case, twisted $(0,2)$ models are not topological.  Due to the absence of the left-moving fermionic symmetry, the cohomology class of an observable varies \emph{holomorphically} in the insertion point (i.e.,~$\del_\zb\CO$ is $Q$-exact if $\CO$ is $Q$-closed).  For this reason, the $Q$-cohomology of local operators will be called the \emph{chiral algebra} of the $(0,2)$ model, which we will denoted by $\CA$.  The non-topological nature of $(0,2)$ models also appears in other aspects of their chiral algebras; for example, they in general receive perturbative corrections,  in addition to nonperturbative ones which are familiar from the $(2,2)$ case.

In this paper, we will consider $(0,2)$ models with K\"ahler target spaces, which contain no left-moving fermions at all.  Such a model has a bosonic field $\phi\colon \Sigma \to X$, and right-moving fermionic fields $\psi_+^i$, $\psi_+^\ib$ which are worldsheet spinors taking values respectively in $TX$ and $\TXb$.  The dynamics of the model is governed by the action
\begin{equation}
S =
\int_\Sigma d^2z \, g_{i\jb}
\bigl(\del_\zb\phi^i \del_z\phi^\jb + \psi_+^i D_z\psi_+^\jb \bigr)
+ \int_\Sigma \phi^*\,\omega,
\end{equation}
where $D_z\psi_+^\jb = \del_z\psi_+^\jb + \del_z\phi^\jb \Gamma^\ib_{\jb\kb}\psi_+^\kb$ is the covariant derivative and $\omega = ig_{i\jb} d\phi^i \wedge d\phi^\jb$ is the K\"ahler form of $X$.  There is essentially only one way to twist the theory; in our convention, $\psi_+$ becomes a worldsheet one-form, while $\psib_+$ becomes a worldsheet scalar.  To make this clear, we will write
\begin{equation}
\rho_\zb^i = \psi_+^i, \qquad
\alpha^\ib = \psi_+^\ib.
\end{equation}
The fermionic symmetry $Q$ acts by
\begin{equation}
\label{Q}
\begin{alignedat}{3}
\delta\phi^i     &=  0, &\qquad
\delta\phi^\ib   &= \alpha^\ib, \\
\delta\rho_\zb^i &= -\del_\zb\phi^i, & \delta\alpha^\ib &=  0.
\end{alignedat}
\end{equation}
Under the right-moving $R$ symmetry, $\alpha$ has charge $1$ and $\rho$ has charge $-1$.  Nonperturbatively, the $R$ symmetry is anomalously broken to $\Z_k$, where $k$ is the greatest divisor of the first Chern class $c_1(X)$.  Note that in perturbation theory, a local operator of negative $R$ charge necessarily has antiholomorphic scaling dimension $\hb > 0$.

At the perturbative level, the chiral algebras of these models were studied by Witten~\cite{W1}.  One of the main results of \cite{W1} is that the chiral algebras can be reconstructed, to all orders in perturbation theory, by gluing \emph{free $\beta\gamma$ CFTs} over the target space.  The obstruction to doing this consistently is encoded in the familiar $p_1(X)$ anomaly and the additional $c_1(\Sigma)c_1(X)$ anomaly introduced by twisting.  This connects to the subject of chiral differential operators previously developed by Malikov et al.~\cite{MSV}.  The description of chiral algebras by free CFTs naturally extends to include left-moving fermions \cite{T}.

In \cite{TY}, the authors studied the nonperturbative effects on the chiral algebra of the model with $X = \CP^1$.  The perturbative chiral algebra of the $\CP^1$ model is a nontrivial affine $\sl_2$-module \cite{MSV, W1}.  Nonperturbatively, however, it becomes trivial (i.e.~identically zero) in the presence of worldsheet instantons.  It then follows that the cohomology of states is also trivial, and therefore, supersymmetry is spontaneously broken in this model.  The cohomology of states can in fact be considered as the cohomology of the Dirac operator on the loop space $LX$, whose index computes the Witten genus of~$X$~\cite{W2, W3}.  The result of \cite{TY} thus indicates that the kernel of the Dirac operator on $LX$ is empty.  This is consistent with Stolz's idea~\cite{S} that if~$X$ has positive Ricci curvature---such is the case for $X = \CP^1$---then the scalar curvature of $LX$ is positive and hence there are no harmonic spinors.

In the following, we will see that the argument that led to the trivialization of the chiral algebra of the $\CP^1$ model actually carries over to a larger class of target spaces.  Particularly interesting examples of such target spaces are complete flag manifolds $G/T$ of compact semisimple Lie groups $G$, where $T$ is the maximal torus.  These include $SU(3)/U(1) \times U(1)$, which was recently found by Tomasiello~\cite{To} to be a novel flux vacuum of string theory.  A similar result, albeit derived via a different approach, has also been obtained by Frenkel et al.~\cite{FLN}.

\section{The Mechanism of Trivialization}

We will begin our discussion with general considerations on a mechanism that renders the chiral algebra trivial.  First, note that the chiral algebra is trivial if and only if there exists a local operator $V$ which satisfies
\begin{equation}
\label{QV1}
\{Q,\Theta\} = 1.
\end{equation}
This is true, for if $\CO$ is a $Q$-closed operator, then $\CO = \{Q, \Theta\CO\}$; conversely, if the $Q$-cohomology is trivial, then the constant operator $1$ must be $Q$-exact since it is $Q$-closed.  A trivial chiral algebra characterized by \eqref{QV1} implies that the $Q$-cohomology of states is also trivial: a $Q$-closed state $\ket{\Psi}$ can be written as $\ket{\Psi} = Q (\Theta \ket{\Psi})$.  Clearly, it suffices for our purpose to find an operator $V$ which gives
\begin{equation}
\label{QVW}
\{Q,V\} = W,
\end{equation}
where $W$ is an \emph{invertible} local operator---by the nilpotency of the supersymmetry transformation, $W$ must be $Q$-closed, and thus $\{Q, W^{-1} V\} = 1$.  The basic observation is that the relation \eqref{QV1} cannot be induced by perturbative effects.  For \eqref{QV1} to hold at the perturbative level, $\Theta$ must have charge~$-1$, since $Q$ has charge $1$.  This implies that~$\Theta$ has $\hb > 0$.  But since the scaling dimension is not violated perturbatively, and $Q$ is dimensionless, $\{Q, \Theta\}$ cannot be equal to~$1$.  Hence, the trivialization of the chiral algebra can only be a purely nonperturbative phenomenon, induced by worldsheet instantons.  It is thus useful to pause here and discuss the structure of instanton corrections.

In the present case, an instanton $\phi_0$ is a holomorphic map from the worldsheet $\Sigma$ to the target space $X$.  We normalize the K\"ahler form of $X$ so that
\begin{equation}
\int_\Sigma \phi_0^*\,\omega = kt,
\end{equation}
where $k \geq 0$ is an integer called the degree of the instanton.  An instanton of degree $k$ will be called a $k$-instanton, and the space of $k$-instantons will be denoted by $\CM_k$.  Zero-instantons are constant maps, whose moduli space is the target space itself: $\CM_0 \iso X$.  The path integral decomposes into  instanton sectors, and the correlation function takes the form
\begin{equation}
\label{CORR}
\bigvev{\dots} = \sum_{k = 0}^\infty e^{-kt} \bigvev{\dots}_k.
\end{equation}
The $k$-instanton correlation function $\vev{\dots}_k$ involves an integration over $\CM_k$.  In particular, the perturbative correlation function $\vev{\dots}_0$ contains an integration over $X$.

An operator relation such as \eqref{QV1} or \eqref{QVW} can be established by comparing correlation functions with arbitrary insertions.  Alternatively, one may compare the matrix elements between arbitrary states.  To obtain the matrix element $\bra{a} A \ket{b}$ of an operator~$A$, one quantizes the theory on a cylinder of infinitesimal length, and compute the corresponding path integral with the initial state $\ket{b}$ and the final state $\ket{a}$.  If the theory is conformally invariant, one can rescale the cylinder to make it infinitely long, and then compactify to a sphere by adding points at infinity.  The matrix element then reduces to a three-point function with vertex operators inserted at $z = 0$ and $\infty$:
\begin{equation}
\label{ME3PTF}
\bra{a} A \ket{b}
= \bigvev{\CV_a(\infty) A \CV_b(0)}.
\end{equation}
Together with \eqref{CORR}, this makes it clear that one can expand $A$ in the instanton weight to obtain
\begin{equation}
\label{OPEXP}
A = \sum_{k=0}^\infty e^{-kt} A_k
\end{equation}
as an operator relation.  Starting with a classically conformally invariant theory, the form of the expansion \eqref{OPEXP} remains the same even when conformal invariance is broken by quantum corrections.  This is because at the lowest order in the perturbation theory around instantons, the theory can be treated as conformally invariant; the explicit computations of correlation functions will be carried out with respect to a free field action that is quadratic in the fluctuating fields%
\footnote{Actually, the path integral measure may transform nontrivially under a conformal transformation, in which case conformal invariance is broken even at the lowest order.  However, in proving a relation such as \eqref{QV0Wl}, one can redefine the $k$-instanton measure by multiplying a quantity $C_k$ that cancels the conformal anomaly.  The theory with the modified measure is now conformally invariant.  To be consistent, one then has to multiply the operators by $C_k^{-1}$.  Note that $W_{l+1}$ on the right-hand side of \eqref{QV0Wl} must be captured by the zeroth instanton sector.  Then, the correct operator relation will be $C_{l+1}^{-1} \{Q,V\} = e^{-(l+1)t} C_0^{-1} W_{l+1} + \dotsm$.  }
Then, the matrix elements can still be computed from three-point functions.  However, $A_k$ will now receive perturbative corrections.

A subtlety in the above argument is that the twisted model which we have been discussing so far is anomalous on the sphere if $c_1(X) \neq 0$.  However, note that on the cylinder or the sphere with two points removed, the canonical line bundle is trivial.  Consequently, the twisting does nothing upon choosing the trivial spin structure, and the physical and twisted models are equivalent.  
As such, one may as well work with the physical model, which is free of the above anomaly anyway.

Let us return to the relation \eqref{QVW}, which characterizes the trivialization of the chiral algebra.  Suppose that the trivialization occurs at the $(l+1)$-instanton level with $l \geq 0$, so that up to the $l$-instanton level, the chiral algebra is nontrivial.  Then, there exist operators $V$ and~$W$ which satisfy
\begin{equation}
\label{QV0Wl}
\{Q,V\} = W =  e^{-(l+1)t} W_{l+1} + \dotsb,
\end{equation}
where $W$ is expanded in accordance with~\eqref{OPEXP}.  This equation implies that $V$ is $Q$-closed up to the $l$-instanton level.  Let us see if $V$ can be $Q$-exact at the $m$-instanton level for any $m \leq l$.  If not, then $V$ will represent a cohomology class up to the $l$-instanton level.  If $V$ were to be $Q$-exact at the $m$-instanton level, then it would be written as $V = \{Q,U\} + e^{-(m+1)t} V'$ for some $U$ and $V'$.  Plugging this into \eqref{QV0Wl}, one finds
\begin{equation}
\label{QV0W2}
\{Q,V'\} = W' = e^{-(l-m)t} W'_{l-m} + \dotsb
\end{equation}
with $W'_{l-m} = W_{l+1}$.  Notice that $W'$ is invertible if and only if $W'_{l-m}$ is invertible, for small higher order corrections cannot affect the invertibility.  Now, we know that $W$ is invertible.  Applying the same argument, we conclude that its lowest order term $W_{l+1} = W'_{l-m}$ is also invertible, and hence so is $W'$.  It then follows from \eqref{QV0W2} that the chiral algebra is trivialized at the $(l-m)$-instanton level.  But since we assumed that the chiral algebra was nontrivial up to the $l$-instanton level, this is a contradiction.  Hence, $V$ must represent a cohomology class up to the $l$-instanton level.

The above argument about the invertibility of $W$ also shows that the chiral algebra will remain trivial through all higher instanton levels if it is trivialized at the $(l+1)$-instanton level.  In this paper, we will present examples where this occurs at $l = 0$, i.e.,~the operator $V$ is a perturbative $Q$-cohomology class.  Essentially, the trivialization will be a consequence of the fact that the grading by the $R$ charge is anomalously broken to $\Z_2$ in all these examples, in a specific way that will be made clear later.

\section{The $\CP^1$ Model Revisited}

As discussed in \cite{TY}, the trivialization of the chiral algebra is realized in the model with target space $X = \CP^1$.  This is the simplest yet most important example, for it serves as a basis for the analysis of more interesting examples such as complete flag manifolds of compact semisimple Lie groups.  As such, it will be useful to first review how the trivialization occurs in the $\CP^1$ model.

We will denote by $\CA^q$ the $Q$-cohomology group of charge $q$.  In the following, by $\CA^\bullet$ we will always mean the perturbative chiral algebra unless otherwise noted.  In general, there exists a cohomology class $\theta \in \CA^1$ associated to $c_1(X)$ whenever it is nonzero.  Explicitly,~$\theta$ is given by~\cite{TY}
\begin{equation}
\theta = R_{i\jb} \del_z\phi^i \alpha^\jb.
\end{equation}
On the other hand, a nonzero $c_1(X)$ breaks conformal invariance, since the one-loop beta function is proportional to the Ricci curvature.  In fact, these two facts are closely related: Although classically $T_{zz}$ is $Q$-closed, perturbatively one has
\begin{equation}
\label{QTDT}
[Q, T_{zz}] = \del_z\theta.
\end{equation}
Thus, perturbative corrections lift $T_{zz}$ out of the classical chiral algebra, and connect it to~$\del_z\theta$.  As a result, the chiral algebra of a model with $c_1(X) \neq 0$ lacks conformal invariance.  Notice that perturbative corrections can only connect a pair of local operators of the same scaling dimensions.  Instantons, however, need not do so.

From \eqref{QTDT}, we find a similar feature between $\theta$ and the constant operator~$1$; namely, both of these operators have vanishing derivatives in the $Q$-cohomology.  In the $\CP^1$ model, there is an even deeper connection between them.  Reflecting the geometry of the target space $\CP^1 = \SL_2/B$, the model possesses affine $\sl_2$ (or $\widehat\sl_2$) currents in $\CA^0$.  The level of the affine algebra is necessarily $-2$, which is the critical level, for otherwise there will be a stress-energy tensor by the Sugawara construction.  The action of these currents by the OPE makes $\CA^0$ and $\CA^1$ naturally an $\widehat{\sl}_2$-module.  In fact, they are isomorphic $\widehat{\sl}_2$-modules, where the isomorphism is given by $\CO \mapsto \CO\theta$ \cite{MSV}.  The perturbative chiral algebra is therefore constructed by acting ``creation operators'' $\CO \in \CA^0$ on the ``ground states'' $1 \in \CA^0$ and $\theta \in \CA^1$.  This is analogous to the Ramond spectrum of strings.

Being a cohomology class in the zero-instanton sector, $\theta$ satisfies the requirement to be an operator responsible for the trivialization of the chiral algebra.  Since $c_1(\CP^1) = 2x$, where $x$ is the generator of $H^2(\CP^1,\Z)$, the charge violation is $2$ in the one-instanton sector and the grading by $R$ charge is broken to $\Z_2$.  Hence, one may expect that there will be a relation
\begin{equation}
\label{QTHETA1}
\{Q, \theta\} \sim e^{-t}
\end{equation}
in the presence of one-instantons.  If such a relation exists, it means that the inverse map of $\CO \mapsto \CO\theta$ is nothing other than the supercharge $Q$.

Although the counting of charge violation works out, there is still a somewhat mysterious property of \eqref{QTHETA1} we have to account for.  Plugging \eqref{QTHETA1} into the correlation function \eqref{CORR}, we find
\begin{equation}
\label{QTHETA12}
\bigvev{\{Q,\theta\} \dotsm}_1 \sim \bigvev{1 \dotsm}_0.
\end{equation}
The left-hand side of this equation involves an integration over $\CM_1$.  The right-hand side, on the other hand, involves an integration over $\CM_0 \iso X$.  Therefore, $\{Q, \theta\}$ must somehow transform the measure of $\CM_1$ into that of $\CM_0$.  How can this happen?

Suppose that we wish to compute the matrix elements of $\{Q,\theta\}$.  We will restrict here to the lowest order in the perturbation theory around instantons, so we may freely exploit conformal field theory arguments.  As discussed in the previous section, an arbitrary matrix element can then be obtained from a three-point function on the sphere:
\begin{equation}
\label{3PTF}
\bra{a} \{Q,\theta\}(z,\zb) \ket{b}
= \bigvev{\CV_a(\infty) \{Q,\theta\}(z,\zb) \CV_b(0)}.
\end{equation}
Let us use a M\"obius transformation on the worldsheet to rearrange the locations of the operators, so that the three-point function \eqref{3PTF} becomes
\begin{equation}
\label{3PTF2}
\bigvev{\{Q',\theta'\}(1) \CV'_a(\epsilon,\epsilonb) \CV'_b(0)}
\end{equation}
for some $|\epsilon| \ll 1$, where the prime indicates that the operators are expressed in this new frame.  We can expand the operators as $\CV'_a = f_a{}^i(\epsilon, \epsilonb) \CA_i$ using a complete set of local operators, where the coefficients $f_a{}^i$ depend on the choice of $\epsilon$.  We then have
\begin{equation}
\label{3PTF3}
f_a{}^i f_b{}^j(\epsilon, \epsilonb) 
\bigvev{\{Q',\theta'\}(1) \CA_i(\epsilon,\epsilonb) \CA_j(0)}.
\end{equation}
At this point, we use the OPE
\begin{equation}
\label{OPE}
\CA_i(\epsilon,\epsilonb) \CA_j(0)
= \sum_k c^k{}_{ij}(\epsilon,\epsilonb) \CA_k(0).
\end{equation}
The OPEs can of course produce short-distance singularities as $\epsilon \to 0$.  Thus if we started with a finite matrix element, $f_a{}^i f_b{}^j(\epsilon, \epsilonb)$ in \eqref{3PTF3} must contain a factor $\epsilon^m \epsilonb^n$ of appropriate powers $m$, $n$ to cancel the most singular terms that appear in the OPEs.  These most singular terms will then be the only terms in the OPEs that contribute to the matrix element in the limit $\epsilon \to 0$.  In particular, these terms arise from the complete contraction of fermionic fields (and their derivatives) and the complete contraction of the derivatives of the bosonic field.  Therefore, the computation ultimately boils down to two-point functions of the form
\begin{equation}
\label{2PTF}
\bigvev{\{Q,\theta\}(1) \CV(0)},
\end{equation}
where $\CV$ is a \emph{function} of the bosonic field.

We can now understand qualitatively how the transmutation of the instanton measure occurs.  We wish to compute the two-point function \eqref{2PTF} in the one-instanton sector.  In the present case, a one-instanton $\phi_0$ is a biholomorphic map from the Riemann sphere to $X = \CP^1$, namely, a M\"obius transformation.  It can be described by three parameters, which we will conveniently take to be the points in $X$ that the points $z = 0$, $1$, and $\infty$ on the worldsheet are mapped to.  The one-instanton computation thus involves integrations over $\phi_0(0)$, $\phi_0(1)$, and $\phi_0(\infty)$.  Schematically, the $\phi_0(\infty)$ integration will give a constant, the $\phi_0(1)$ integration will give an integration of $\{Q,\theta\}$ which will again be a constant.  Finally, the $\phi_0(0)$ integration will become an integration of the function $\CV$ over the target space.  But this is just the one-point function $\vev{\CV(0)}_0$ in the zero-instanton sector.  Consequently, we will have a relation
\begin{equation}
\bigvev{\{Q,\theta\}(1) \CV(0)}_1 \sim \bigvev{\CV(0)}_0,
\end{equation}
which is equivalent to \eqref{QTHETA12}.

In the rest of this section, we will make the above argument more precise and explicit.  Let us first find the number of fermionic zero modes.  Recall from the discussion in the previous section that here we are dealing with the physical model, so the fermions are spinors.  The $\alpha$ and $\rho$ zero modes obey
\begin{equation}
\del_z \alpha^\ib = \del_z \rho_\ib = 0.
\end{equation}
Hence, after taking the complex conjugate, they are respectively holomorphic sections of $\CO(-1) \otimes \phi_0^*TX$ and $\CO(-1) \otimes \phi_0^*T^*X$, where $\CO(-1)$ is the spin bundle.  For one-instantons, $\phi_0^*TX = \CO(2)$ and $\phi_0^*T^*X = \CO(-2)$.  Since $h^0(\CO(1)) = 2$ and $h^0(\CO(-3)) = 0$, we have two $\alpha$ zero modes, and no $\rho$ zero modes.  We see that $\{Q,\theta\}$ indeed contains just the right number of $\alpha$~fields to soak up the fermionic zero modes.%
\footnote{If there exists more fermionic zero modes, we must bring down a term proportional to the Riemann curvature from the action.  This contains one $\alpha$ and one $\rho$ zero mode, and contributes to the lowest order in the large volume limit.}

In computing $\{Q,\theta\}$, one must look for antiholomorphic single poles in the OPE of~$J(\zb) \cdot \theta(w,\wb)$, where $J = g_{i\jb} \del_\zb\phi^i \alpha^\jb$ is the supercurrent and $Q = \oint J d\zb$.  This OPE contains two~$\alpha$ fields.  Notice that the Taylor expansion of $\alpha(\zb) \alpha(\wb)$ around $\wb$ starts with the first order in $\zb - \wb$, since the zeroth order vanishes by the grassmannian nature of $\alpha$.  Thus, in order to get a single pole in the overall computation, we must look for double poles in the contractions of the rest of the fields in $J$ and $\theta$.  In view of the expression $\theta = R_{i\jb} \del_z\phi^i\alpha^\jb$, it is not at all obvious how such poles arise.  The crucial observation made in \cite{TY} is that $\alpha$ is a section of $\phi^*\TXb$, and here $\phi$ is not constrained to instantons.  In other word, the fermions carry within themselves fluctuating bosonic nonzero modes.  It turns out \cite{TY} that~$\theta$ in the one-instanton sector can be written as $\theta = \theta' + \dotsb$, where
\begin{equation}
\label{THETA'}
\theta'
= R_{i\jb}(\phi_0,\phib_0) \del_z\phi_0^i
  \frac{\del_\zb\phi^\jb}{\del_\zb\phi_0^\kb} \alpha_0^\kb.
\end{equation}
Here, $\alpha_0$ is the zero mode part of $\alpha$ evaluated at $\phi_0$.  The contraction between $\del_\zb\phi^i$ in~$J$ and $\del_\zb\phi^\jb$ in~$\theta'$ can now produce an antiholomorphic double pole.  After the contour integration,  this yields 
\begin{equation}
\label{QTHETA'}
\{Q, \theta'\}
= R_{i\jb}(\phi_0,\phib_0) \frac{\del_z\phi_0^i}{\del_\zb\phi_0^\kb}
  \del_\zb\alpha_0^\jb \alpha_0^\kb,
\end{equation}
up to irrelevant terms that are of higher orders in the perturbation theory around instantons.  The other terms in $\theta$ contribute to $\{Q,\theta\}$ obviously as $Q$-exact terms.  Hence, they can be ignored.

To complete the computation, we need the explicit forms of the fermionic zero modes.  If we express the one-instantons as
\begin{equation}
\label{MOEBIUS}
\phi_0(z) = \frac{az + b}{cz + d}; \qquad
ad - bc = 1,
\end{equation}
then $\alpha_0$ can be expanded as $\alpha_0^\ib = c_0^1 \ub_{0,1} + c_0^2 \ub_{0,2}$ with grassmannian coefficients $c_0^1$, $c_0^2$, where $\ub_{0,1}$, $\ub_{0,2}$ are \cite{DG}
\begin{equation}
\ub_{0,1}(\zb) = \frac{1}{\bar c\zb + \bar d}, \qquad
\ub_{0,2}(\zb) = \frac{1}{\bar c(\bar c\zb + \bar d)^2}.
\end{equation}
After the fermionic zero mode integration, $\{Q,\theta'\}$ becomes the pullback of the K\"ahler form:
\begin{equation}
\label{PBKAHLER}
\int \! dc_0^1 dc_0^2 \, \{Q,\theta'\}
= R_{i\jb} \del_z\phi_0^i \del_\zb\phi_0^\jb.
\end{equation}
This must be plugged into the two-point function \eqref{2PTF}, and then integrated over the one-instanton sector.  In the region $d \neq 0$, one can set $d = 1$ by an overall rescaling.  The conformally invariant measure on $\CM_1$ is, up to an overall constant, given by \cite{DSWW}
\begin{equation}
\label{DM1}
d\CM_1 = |ad - bc|^{-4} \, d ^2a \, d^2b \, d^2c.
\end{equation}
Recalling that $\CV$ is a function, the two-point function \eqref{2PTF} can be evaluated as
\begin{equation}
\int \frac{d^2Y}{|Y|^4}
\int \! d^2X_1 \, R_{1\bar 1}(X_1, \Xb_1)
\int \! d^2X_0 \, \CV(X_0, \Xb_0).
\end{equation}
Here, $X_0 = \phi_0(0) = b$, \ $X_1 = \phi_0(1) = (a + b)/(c + 1)$, \ $Y = a - bc$.   The first divergent integral in $Y$ reflects the noncompactness of the M\"obius group.  The second integral comes from $\{Q,\theta\}$, and is the integration of $c_1(X)$.  And the third integral is an integration of $\CV$ over the target space, which will give the zero-instanton one-point function $\vev{\CV(0)}_0$.

More precisely, let $d\CM_0 = \Omega(X_0) d^2X_0$ be the volume form of the zero-instanton sector.  The nowhere vanishing function $\Omega$ defines an invertible operator in the zero-instanton sector through its matrix elements $\Omega(X_0)$ (expressed in the basis where states are localized in the target space).  The computation above demonstrates that we have the operator relation
\begin{equation}
\label{QTHETA1PRECISE}
\{Q,\theta\} \sim e^{-t} \Omega^{-1}.
\end{equation}
We have found a relation of the form \eqref{QV0Wl}, which via \eqref{QVW} implies that we have the relation $\{Q,\Theta\} = 1$.  Therefore, the chiral algebra is trivialized in the $\CP^1$ model.

\section{Higher Dimensional Target Spaces}

The key property of the $\CP^1$ model that was crucial for the trivialization of its chiral algebra is that there are precisely two $\alpha$ zero modes and no $\rho$ zero modes in the one-instanton sector.  In this case a perturbative cohomology class $\theta$ exists (since $c_1(\CP^1) \neq 0$), and $\{Q,\theta\}$ contains the right number of fermionic zero modes.  Hence, any one-instanton correlation function with $\{Q,\theta\}$ inside reduces to a zero-instanton one, and this establishes the relation \eqref{QTHETA1PRECISE}.  Let us see if $\{Q,\theta\}$ leads to a similar relation in the case of other nonanomalous target spaces that have $p_1(x) = 0$ but nonvanishing $c_1(X)$.

Let $X$ be a K\"ahler manifold of complex dimension $d$, and $\phi_0$ a one-instanton wrapping a rational curve $L \subset X$.  We can decompose the tangent bundle as $TX = TL \oplus NL$, where $NL$ is the normal bundle of $L$ in $X$.  The pullback bundle $\phi_0^*TL = \CO(2)$ for one-instantons, and $\phi_0^*NL$ further splits into the direct sum of line bundles.  If $\int_\Sigma \phi_0^*c_1(X) = k$, then we can write
\begin{equation}
\label{ST}
\phi_0^*TX
\iso \CO(2) \oplus \CO(p_1) \oplus \dotsb \oplus \CO(p_{d-1}),
\end{equation}
where $p_1 + \dotsb + p_{d-1} = k - 2$.  The number of $\alpha$ or $\rho$ zero modes can be found from the splitting type \eqref{ST}.  According to the formula
\begin{equation}
h^0\bigl(\CO(n)\bigr) =
\begin{cases}
n + 1 & \text{for $n \geq 0$}; \\
0     & \text{for $n < 0$},
\end{cases}
\end{equation}
each $\CO(n)$ with $n > 0$ in \eqref{ST} contributes $n$ of $\alpha$ zero modes.  We also know from the index theorem that there are $k$ more $\alpha$ zero modes than $\rho$ zero modes.  In order to have exactly two $\alpha$ and no $\rho$ zero modes, it must be that $k = 2$ and the splitting type is
\begin{equation}
\label{2000}
\CO(2) \oplus \CO(0) \oplus \dotsb \oplus \CO(0).
\end{equation}
Notice then that the two $\alpha$ zero modes should come solely from $TL$.  This means that only the field components tangent to~$L$ contribute to $\{Q,\theta\}$.  Consequently, our computation will be the same as that in the $\CP^1$ case, and the fermionic zero mode integration turns $\{Q,\theta\}$ into the pullback of the K\"ahler form~\eqref{PBKAHLER}, but this time restricted to $L$.  The integration over the parameters of the instanton then becomes an integration over $L$.

However, this is not quite the end of the story, since $L$ is not a rigid instanton if \eqref{2000} is true.  An infinitesimal deformation of $L$ is given by a holomorphic section of~$\phi_0^*NL$.  In the case of the splitting type~\eqref{2000}, we have $d - 1$ independent deformations, one for each normal direction.  Intuitively, we therefore expect that the instanton can be infinitesimally translated in every possible direction in the target space.  This generates a family of instantons with $d-1$ complex parameters, over which we still have to integrate after the integration over $L$ is done.  If the instanton sweeps the whole target space, then we will obtain an integration over the target space.

This can happen for homogeneous spaces $G/H$ equipped with a $G$-invariant K\"ahler structure.  Assuming that the model with target space $G/H$ has a number of topologically distinct one-instantons, pick one and call it $\phi_0$.  Then, the transitive $G$-action can map $\phi_0$ to another instanton located anywhere else in $G/H$.  This means that there is at least one deformation in every normal direction, i.e., all the $p_i$'s in the splitting type \eqref{ST} must be nonnegative.  If any of the $p_i$'s is positive, then the instanton has charge violation greater than $2$, and hence, does not contribute to $\{Q,\theta\}$.  Thus, in order for the chiral algebra to be trivialized, it suffices that there exists at least a single one-instanton such that the tangent bundle has splitting type \eqref{2000}.

This is indeed the case for flag manifolds $G/T$ of compact semisimple Lie groups~$G$!  It is well-known \cite{BH} that $c_1(G/T) = 2(x_1 + \dots + x_r)$, where $r$ is the rank of $G$.  Furthermore, there are $r$ rational curves that are dual to the $x_i$'s.  Each of the rational curves (i.e., instantons) has the splitting type \eqref{2000} by the above argument.  Therefore, we conclude that the chiral algebras for complete flag manifolds are zero nonperturbatively.

\section{Loop Space Geometry}

We have seen that the chiral algebras of $(0,2)$ models are trivialized by instantons for a large class of target spaces given by flag manifolds.  Note that supersymmetric states must be annihilated by $Q$ and $Q^\dagger$, hence they are harmonic states of $Q$.  Since the harmonic space is isomorphic to the $Q$-cohomology of states, supersymmetry is spontaneously broken in the models with these target spaces.  As we now explain, this is also echoed in the geometry of their loop spaces.

To unravel the connection between supersymmetry and the loop space geometry, let us consider the model with target space $X$ defined on the cylinder $S^1 \times \R$, and regard it as supersymmetric quantum mechanics on the loop space $LX$.  By canonical quantization, one obtains the anticommutation relation
\begin{equation}
\label{CAR}
\{\rho_\ib(\sigma,\tau), \alpha^\jb(\sigma',\tau)\}
= \delta_\ab^\bb \delta(\sigma - \sigma'),
\end{equation}
where $(\sigma,\tau)$ are the coordinates on the cylinder.  This is the loop-space analog of the Clifford algebra.  The supercharge is identified as
\begin{equation}
\label{Q}
Q  = \int_{S^1} \! d\sigma 
     \Bigl(\alpha^\ib \frac{D}{D\phi^\ib}
     - i g_{\ib j} \alpha^\ib \del_\sigma\phi^j\Bigr),
\end{equation}
where $D/D\phi^\ib$ is the covariant functional derivative on $LX$.  Notice that the first term of~\eqref{Q} is just (a half of) the Dirac operator on $LX$.  In fact, the supercharge $Q$ is related to the Dirac operator by a similarity transformation~\cite{TY}, which does not affect the cohomology.  The $Q$-cohomology of states is therefore the spinor cohomology on $LX$.

In the large volume limit, the theory can also be formulated from the target space perspective.  In this case, perturbative supersymmetric states take the form \cite{W2}
\begin{equation}
\label{STATE}
f_{i_1 \dotsm i_k}{}^{j_1 \dotsm j_l}
c_{-m_1}^{i_1} \dotsm c_{-m_k}^{i_k}
c_{-n_1,j_1} \dotsm c_{-n_l,j_l} \ket{\psi},
\end{equation}
where $\ket{\psi}$ is a spinor ground state, and $c_{-m}^i$, $c_{-n,i}$ are left-moving bosonic creation operators.  The state \eqref{STATE} can be viewed as a section of the twisted spin bundle $S \otimes R_m \otimes R_n^*$, where $m = m_1 + \dotsb + m_k$, \  $n = n_1 + \dotsb + n_l$, and
\begin{equation}
\label{REP}
\sum_{k = 0}^\infty q^k R_k
= \bigotimes_{k = 0}^\infty \bigoplus_{l = 0}^\infty
  \Sym^l(q^k \cdot T^*X).
\end{equation}
The supercharge $Q$, when acting on these states, reduces to the Dirac operator on $X$ twisted by the relevant bundle given by \eqref{REP}.

Our results thus imply the following.  From the loop space viewpoint, the spinor cohomology on $LX$ is zero for $X = G/T$, i.e., complete flag manifolds of compact semisimple Lie groups.  Equivalently, this means that the kernel of the Dirac operator on $LX$ is empty.  This is consistent with Stolz's idea~\cite{S} that if~$X$ has positive Ricci curvature, then $LX$ has positive scalar curvature with no harmonic spinors; flag manifolds indeed admit positive Ricci curvature~\cite{B}.  From the target space viewpoint, the kernels of the Dirac operators twisted by \eqref{REP} become empty nonperturbatively.  This can be interpreted as an effect arising from a ``quantum'' deformation of the geometry of~$X$ by worldsheet instantons.

\section*{Acknowledgement}

The authors would like to thank M.~Brion, H.-B. Duan, J.~Fine, T.~H\"ubsch, M.~Nakagawa, S.~Salamon, A.~Tomasiello, and E.~Witten for helpful discussions. The work of M.-C.T. is supported by the Institute for Advanced Study and the NUS -- Overseas Postdoctoral Fellowship.

\end{document}